\documentstyle[12pt]{article}
\catcode`\@=11

\@addtoreset{equation}{section}
\def\theequation{\arabic{section}.\arabic{equation}}

\catcode`\@=11

\def\section{\@startsection{section}{1}{\z@}{3.5ex plus 1ex minus
   .2ex}{2.3ex plus .2ex}{\large\bf}}

\newskip\humongous \humongous=0pt plus 1000pt minus 1000pt

\newif\ifdtup

\def\eqnarray{\let\@currentlabel=\theequation\refstepcounter{equation}
    \global\@eqnswtrue
    \global\@eqcnt\z@\tabskip\@centering\let\\=\@eqncr
    $$\halign to \displaywidth\bgroup\@eqnsel\hskip\@centering
      $\displaystyle\tabskip\z@{##}$&\global\@eqcnt\@ne
       \hfil${{}##{}}$\hfil
      &\global\@eqcnt\tw@ $\displaystyle\tabskip\z@{##}$\hfil
       \tabskip\@centering&\llap{##}\tabskip\z@\cr}
\def\lefteqn#1{\hbox to 4\arraycolsep{$\displaystyle #1$\hss}}
\def\thesection{\arabic{section}.}

\def\appendix{\setcounter{section}{0}
        \def\thesection{Appendix.}
        \def\theequation{\Alph{section}.\arabic{equation}}}

\long\def\@makefntext#1{\parindent 0cm\noindent
\hbox to 1em{\hss$^{\@thefnmark}$}#1}
\def\IR{{\hbox{{\rm I}\kern-.2em\hbox{\rm R}}}}
\def\IH{{\hbox{{\rm I}\kern-.2em\hbox{\rm H}}}}
\def\IC{{\ \hbox{{\rm I}\kern-.6em\hbox{\bf C}}}}
\def\IZ{{\hbox{{\rm Z}\kern-.4em\hbox{\rm Z}}}}
\def\rref#1{(\ref{#1})}
\def\Tr{\hbox{Tr}}
\newcommand{\beq}{\begin{equation}}
\newcommand{\eeq}{\end{equation}}
\newcommand{\NPB}[1]{{\sl Nucl.~Phys.}~{\bf B#1}}
\newcommand{\Ann}[1]{{\sl Ann.~Phys.}~{\bf #1}}
\newcommand{\CMP}[1]{{\sl Commun.~Math.~Phys.}~{\bf #1}}

\newcommand{\PRL}[1]{{\sl Phys.~Rev.~Lett.}~{\bf #1}}

\newcommand{\CQG}[1]{{\sl Class.~Quant.~Grav.}~{\bf #1}}
\newcommand{\PRD}[1]{{\sl Phys.~Rev.}~{\bf D#1}}

\begin{document}
%
%
%
%
\def\citen#1{%
\edef\@tempa{\@ignspaftercomma,#1, \@end, }
\edef\@tempa{\expandafter\@ignendcommas\@tempa\@end}%
\if@filesw \immediate \write \@auxout {\string \citation {\@tempa}}\fi
\@tempcntb\m@ne \let\@h@ld\relax \let\@citea\@empty
\@for \@citeb:=\@tempa\do {\@cmpresscites}%
\@h@ld}
%
\def\@ignspaftercomma#1, {\ifx\@end#1\@empty\else
   #1,\expandafter\@ignspaftercomma\fi}
\def\@ignendcommas,#1,\@end{#1}
%
%
\def\@cmpresscites{%
 \expandafter\let \expandafter\@B@citeB \csname b@\@citeb \endcsname
 \ifx\@B@citeB\relax 
    \@h@ld\@citea\@tempcntb\m@ne{\bf ?}%
    \@warning {Citation `\@citeb ' on page \thepage \space undefined}%
 \else
    \@tempcnta\@tempcntb \advance\@tempcnta\@ne
    \setbox\z@\hbox\bgroup 
    \ifnum\z@<0\@B@citeB \relax
       \egroup \@tempcntb\@B@citeB \relax
       \else \egroup \@tempcntb\m@ne \fi
    \ifnum\@tempcnta=\@tempcntb 
       \ifx\@h@ld\relax 
          \edef \@h@ld{\@citea\@B@citeB}%
       \else 
          \edef\@h@ld{\hbox{--}\penalty\@highpenalty \@B@citeB}%
       \fi
    \else   
       \@h@ld \@citea \@B@citeB \let\@h@ld\relax
 \fi\fi%
 \let\@citea\@citepunct
}
%
\def\@citepunct{,\penalty\@highpenalty\hskip.13em plus.1em minus.1em}%
%
%
\def\@citex[#1]#2{\@cite{\citen{#2}}{#1}}%
%
%
\def\@cite#1#2{\leavevmode\unskip
  \ifnum\lastpenalty=\z@ \penalty\@highpenalty \fi 
  \ [{\multiply\@highpenalty 3 #1
      \if@tempswa,\penalty\@highpenalty\ #2\fi 
    }]\spacefactor\@m}
\let\nocitecount\relax  
%
\begin{titlepage}
\vspace{.5in}
\begin{flushright}
IASSNS-HEP-94/34\\
UCD-94-13\\
gr-qc/9405070\\
May 1994\\
\end{flushright}
\begin{center}
{\Large\bf Aspects of\\[.5ex] Black Hole Quantum Mechanics\\[.5ex]
and Thermodynamics\\[1ex] in 2+1 Dimensions}\\[.5ex]
\vspace{.4in}
S{\sc teven}~C{\sc arlip}\footnote{\it email: carlip@dirac.ucdavis.edu}\\
       {\small\it Department of Physics}\\
       {\small\it University of California}\\
       {\small\it Davis, CA 95616}\\{\small\it USA}\\
\vspace{1ex}
{\small and}\\
\vspace{1ex}
C{\sc laudio}~T{\sc eitelboim}\footnote{\it email: cecsphy@lascar.puc.cl}\\
       {\small\it Centro de Estudios Cientificos de Santiago}\\
       {\small\it Casilla 16443, Santiago 9}\\{\small\it Chile}\\
       {\small\it and}\\
       {\small\it Institute for Advanced Study}\\
       {\small\it Olden Lane, Princeton, NJ 08540}\\{\small\it USA}\\
\end{center}
\begin{center}
\begin{minipage}{5in}
\begin{center}
{\large\bf Abstract}
\end{center}
{\small
We discuss the quantum mechanics and thermodynamics of the (2+1)-dimensional
black hole, using both minisuperspace methods and exact results from
Chern-Simons theory.  In particular, we evaluate the first quantum
correction to the black hole entropy.  We show that the dynamical
variables of the black hole arise from the possibility of a deficit angle
at the (Euclidean) horizon, and briefly speculate as to how they may provide
a basis for a statistical picture of black hole thermodynamics.
}
\end{minipage}
\end{center}
\end{titlepage}
\addtocounter{footnote}{-2}

In the twenty years since Bekenstein's proposal that black holes have
an entropy \cite{Bekenstein} and Hawking's discovery that they can
evaporate \cite{Hawking}, a great deal has been learned about the
thermodynamics of black holes.  Nevertheless, some key questions
remain unanswered:
\begin{enumerate}
\item  Despite considerable work over the past few years, the ``information
loss paradox''---the apparently nonunitary transition from pure particle
states to thermal Hawking radiation---remains an open problem (see, for
example, \cite{Page}).
\item Standard approaches to black hole thermodynamics involve semiclassical
approximations of one kind or another, and can tell us little about the
final stages of the process of evaporation, where the effects of quantum
gravity are sure to be important.
\item Although a number of interesting suggestions have been made, we do
not yet have a generally accepted model of the microscopic statistical
mechanics that should presumably underlie black hole thermodynamics.
\end{enumerate}

The recent discovery of black hole solutions in (2+1)-dimensional gravity
offers a promising new arena for investigating such problems.  In
contrast to (3+1)-dimensional general relativity, the (2+1)-dimensional
model has only finitely many physical degrees of freedom.  As a result,
questions about quantum gravity can be explored in considerable detail,
and we can be reasonably confident that our conclusions are at least
self-consistent.  The purpose of this paper is to begin that exploration.

The plan of the paper is the following.  In section 1, we discuss the
geometry of the Euclidean black hole and its relation to hyperbolic
three-space $\IH^3$, the complete Riemannian space of constant
negative curvature, with appropriate identifications.  We explain
how to generalize these identifications to allow for a conical
singularity (and a helical twist) at the horizon.  The holonomies of
this generalized geometry become the dynamical variables of the black
hole, and are assigned Poisson brackets from previously available results
for the Chern-Simons formulation.  In section 2 we discuss the black hole
from the Hamiltonian point of view, concentrating on a minisuperspace
model.  We again find that the parameters describing a singularity at
the horizon become the dynamical variables of the black hole, with
Poisson brackets equivalent to those derived from the Chern-Simons
approach.  We show that the partition function arises from a sum over
these parameters, and we calculated it in the classical approximation,
obtaining the standard black hole entropy.  We then evaluate the first
quantum correction, again using known results from the Chern-Simons
formulation.  For large black holes, we find that this correction does
not involve $\hbar$, and merely renormalizes the gravitational constant,
a typical occurrence in Chern-Simons theory.  Finally, we devote section
3 to speculation on the possible microscopic origin of black hole entropy
as it emerges from the descriptions obtained in the first two sections.

\section{The Euclidean Black Hole}

We shall investigate black hole thermodynamics in terms of the
``Wick-rotated'' Euclidean black hole.  One may take the point of view
that the Euclidean geometry emerges as a complex stationary point of the
Lorentzian action (see, for example, \cite{BrownYork}), or one may argue
that the Euclidean description is the more fundamental one.  The analysis
that follows applies in either case.

\subsection{Geometry and Identifications}

We start with the static Lorentzian black hole metric \cite{BTZ},
\beq
ds_{\hbox{\scriptsize Lor}}^2 = -\left( {r^2\over\ell^2}-M\right) dt^2
  + \left( {r^2\over\ell^2}-M\right)^{-1} dr^2 + r^2d\phi^2,
\label{a1}
\eeq
which is a solution of the vacuum Einstein equations in 2+1 dimensions
with a cosmological constant $\Lambda = -1/\ell^2$.  This metric and
its spinning counterpart (equation \rref{a8} below) are discussed in
detail in \cite{BHTZ}.  For us, a key feature is the geometric
significance of the time coordinate $t$ as the ``Killing time,'' the
displacement along the timelike Killing vector at spatial infinity.
Since it is this Killing vector that determines the static nature of
the solution and the applicability of equilibrium thermodynamics, $t$
is the appropriate time for thermodynamic considerations.

The metric \rref{a1} has an obvious analytic continuation to
\beq
ds^2 = \left( {r^2\over\ell^2}-M\right) d\tau^2
  + \left( {r^2\over\ell^2}-M\right)^{-1} dr^2 + r^2d\phi^2,
\label{a2}
\eeq
which satisfies the Euclidean field equations.  To analyze the resulting
geometry, we exploit the simplicity of (2+1)-dimensional gravity:
the full curvature tensor in 2+1 dimensions depends linearly on the Ricci
tensor, and any solution of the empty space field equations is a space
of constant curvature.  We can exhibit this characteristic explicitly for
the metric \rref{a2} by changing to coordinates
\begin{eqnarray}
  x &=& \left(1-{M\ell^2\over r^2}\right)^{1/2}\cos{\sqrt{M}\tau\over\ell}\,
       e^{\sqrt{M}\phi} \nonumber\\
  y &=& \left(1-{M\ell^2\over r^2}\right)^{1/2}\sin{\sqrt{M}\tau\over\ell}\,
       e^{\sqrt{M}\phi} \\
  z &=& {\sqrt{M}\ell\over r}e^{\sqrt{M}\phi} . \nonumber
\label{a3}
\end{eqnarray}
The metric becomes
\beq
ds^2 = {\ell^2\over z^2}(dx^2+dy^2+dz^2) ,\quad z>0 ,
\label{a4}
\eeq
which may be recognized as the standard metric for the upper half-space
model of hyperbolic three-space $\IH^3$.  To account for the periodicity
of the Schwarzschild angular coordinate $\phi$, we must make the (isometric)
identifications
\beq
(x,y,z)\sim (e^{2\pi\sqrt{M}}x,e^{2\pi\sqrt{M}}y,e^{2\pi\sqrt{M}}z) .
\label{a5}
\eeq
The (2+1)-dimensional Euclidean black hole may thus be described as the
quotient of hyperbolic space $\IH^3$ by the isometry \rref{a5}.  As in
four dimensions, the Euclidean black hole corresponds to the region outside
the event horizon of the Lorentzian solution; the event horizon itself is
mapped to the circle $x=y=0$, $1\le z\le e^{2\pi\sqrt{M}}$.

It is straightforward to find a fundamental region for the identifications
\rref{a5}.  The result is most easily expressed in ``spherical'' coordinates
\begin{eqnarray}
x &=& R\cos\theta\cos\chi \nonumber\\
y &=& R\sin\theta\cos\chi \\
z &=& R\sin\chi \nonumber
\label{a6}
\end{eqnarray}
with $\theta$ periodic, i.e.,
\beq
(R,\theta,\chi)\sim (R,\theta+2\pi,\chi) .
\label{aa6}
\eeq
The identifications \rref{a5} then become
\beq
(R,\theta,\chi)\sim (e^{2\pi\sqrt{M}}R,\theta,\chi) .
\label{a6b}
\eeq
We may therefore choose as a fundamental region the space between the
hemispheres $R=1$ and $R=e^{2\pi\sqrt{M}}$, with points on the boundaries
identified along radial lines as in figure 1.

Topologically, the resulting manifold is a solid torus.  For
$\chi\ne\pi/2$, each slice of fixed $\chi$ is an ordinary two-torus,
with circumferences parametrized by the periodic coordinates $\ln\!R$
and $\theta$; the degenerate surface $\chi=\pi/2$ is a circle at the
core of the solid torus.  Physically, $R$ is an angular coordinate, equal
to $e^{\sqrt{M}\phi}$ in the original Schwarzschild coordinates; the
azimuthal angle $\theta$ measures time; and $\chi$ is a radial coordinate.
Note that with the metric \rref{a4}, the $x$-$y$ plane $\chi=0$ is
infinitely far from the interior of the manifold, while $\chi=\pi/2$ is
the horizon.

(We caution the reader that it is important to distinguish between the
angular coordinate $\theta$ in the upper half-space representation, which
is physically a time coordinate, and the Schwarzschild angular coordinate
$\phi$.  In the discussion that follows, we shall move freely between these
two useful coordinate systems.)

The metric \rref{a4}--\rref{a5} already gives us important information
about (2+1)-dimensional black hole thermodynamics.  The periodicity
in $\theta$ reflects a periodicity in Killing time, and as usual in
Euclidean quantum field theory, we can interpret the period as an inverse
temperature.  Indeed, comparing \rref{a3} and \rref{a6}, we see that
\beq
\theta = {\sqrt{M}\over\ell}\tau,
\label{a6a}
\eeq
so a period of $2\pi$ in $\theta$ corresponds to a period of
\beq
\beta = {2\pi\ell\over\sqrt{M}}
\label{a7}
\eeq
in the Killing time.

As we shall see in the next section, the off-shell extension of the black
hole solution requires a generalization of this periodicity.  If $\theta$
has a period $\Theta$---that is, if \rref{aa6} is replaced by
\beq
(R,\theta,\chi)\sim (R,\theta + \Theta,\chi)
\label{a7a}
\eeq
---then the metric \rref{a4} acquires a conical singularity along the $z$
axis with deficit angle $2\pi - \Theta$, and will not solve the equations
of motion there.  This mildly singular geometry can also be expressed in
the Schwarzschild coordinates \rref{a2}, where $\Theta$ is now determined
by the condition
\beq
(\tau_2 - \tau_1) (N^\perp)^2 = 2\Theta(r-r_+) + O(r-r_+)^2
\label{a7b}
\eeq
with
\beq
(N^\perp)^2 = {r^2\over\ell^2} -M , \qquad r_+ = \sqrt{M}\ell .
\label{a7c}
\eeq
Equation \rref{a7b} holds for the off-shell black hole in any spacetime
dimension \cite{offshell}.

The extension of this analysis to the spinning black hole is fairly
straighforward.  The Lorentzian metric is now
\beq
ds_{\hbox{\scriptsize Lor}}^2 = -\bigl( N_{\hbox{\scriptsize Lor}}^\perp
  \bigr)^2dt^2 + f_{\hbox{\scriptsize Lor}}^{-2}dr^2
  + r^2\left( d\phi + N^\phi_{\hbox{\scriptsize Lor}} dt\right)^2
\label{a8}
\eeq
with
\beq
N^\perp_{\hbox{\scriptsize Lor}} = f_{\hbox{\scriptsize Lor}}
  = \left( -M_{\hbox{\scriptsize Lor}} + {r^2\over\ell^2}
  + {J_{\hbox{\scriptsize Lor}}^2\over4r^2} \right)^{1/2} ,
  \qquad N_{\hbox{\scriptsize Lor}}^\phi
  = - {J_{\hbox{\scriptsize Lor}}\over2r^2} .
\label{a9}
\eeq
The corresponding Euclidean solution is obtained by letting $t=-i\tau$,
$M_{\hbox{\scriptsize Lor}}=M$, and $J_{\hbox{\scriptsize Lor}}
= iJ$, and reads
\beq
ds^2 = (N^\perp)^2 d\tau^2 + f^{-2}dr^2
 + r^2\left( d\phi + N^\phi d\tau\right)^2
\label{a9a}
\eeq
with
\beq
N^\perp = f = \left( -M + {r^2\over\ell^2} - {J^2\over4r^2}\right)^{1/2} ,
  \qquad N^\phi = -iN^\phi_{\hbox{\scriptsize Lor}} = - {J\over2r^2} .
\label{a9b}
\eeq
If we set
\beq
r_{\pm}{}^2 = {M\ell^2\over2}\left[ 1\pm
  \left(1+{J^2\over M^2\ell^2}\right)^{1/2} \right] ,
\label{a10}
\eeq
it is not hard to check that the coordinate transformation
\begin{eqnarray}
x &=& \left({r^2-r_+^2\over r^2-r_-^2}\right)^{1/2}
      \cos\left( {r_+\over\ell^2}\tau + {|r_-|\over\ell}\phi \right)
      \exp\left\{ {r_+\over\ell}\phi - {|r_-|\over\ell^2}\tau \right\}
      \nonumber \\
y &=& \left({r^2-r_+^2\over r^2-r_-^2}\right)^{1/2}
      \sin\left( {r_+\over\ell^2}\tau + {|r_-|\over\ell}\phi \right)
      \exp\left\{ {r_+\over\ell}\phi - {|r_-|\over\ell^2}\tau \right\} \\
z &=& \left({r_+^2-r_-^2\over r^2-r_-^2}\right)^{1/2}
      \exp\left\{ {r_+\over\ell}\phi - {|r_-|\over\ell^2}\tau \right\}
      \nonumber
\label{a11}
\end{eqnarray}
again takes the metric to the form \rref{a4}.  Here we have set
\beq
|r_-| = ir_- = {J\ell\over2r_+}.
\label{a11a}
\eeq
The identifications analogous to those of to equation \rref{a5} or
\rref{a6b}, in ``spherical'' coordinates \rref{a6}, are now
\beq
(R,\theta,\chi) \sim
  (Re^{2\pi r_+/\ell}, \theta + {2\pi|r_-|\over\ell}, \chi) ,
\label{a12}
\eeq
reducing to \rref{a6b} when $J=0$.  The fundamental region is again the
region between two hemispheres, as in figure 1, but the boundaries are
now identified with a twist around the $z$ axis.

In terms of the Schwarzschild coordinates $\phi$ and $\tau$, the
identification \rref{a12} simply reads
\beq
(\phi,\tau) \sim (\phi+2\pi,\tau) ,
\label{a12a}
\eeq
whereas the demand that $\theta$ should have a period $2\pi$---that
is, the absence of a conical singularity in the Euclidean black hole
metric---translates into
\beq
(\phi,\tau) \sim (\phi+\Phi,\tau+\beta)
\label{a12b}
\eeq
with
\beq
\beta = {2\pi r_+\ell^2\over r_+^2 - r_-^2} ,\qquad
\Phi =  {2\pi |r_-|\ell\over r_+^2 - r_-^2} .
\label{a13}
\eeq
Equation \rref{a13} shows that the angle $\phi$ appearing in \rref{a9a}
is not the usual Schwarzschild azimuthal angle, for which the identification
\rref{a12b} would read
\beq
(\phi',\tau) \sim (\phi',\tau+\beta) .
\label{a13a}
\eeq
Rather, the relationship between $\phi$ and $\phi'$ is evidently
\beq
\phi' = \phi - {\Phi\over\beta}\tau = \phi - {|r_-|\over\ell r_+}\tau .
\label{a13b}
\eeq
It follows that
\beq
d\phi' + N^{\phi'}d\tau = d\phi + N^\phi d\tau
\label{a13c}
\eeq
with
\beq
N^{\phi'} = N^\phi + {|r_-|\over\ell r_+}
  = {|r_-|\over\ell r_+} - {r_+|r_-|\over\ell r^2} ,
\label{a13d}
\eeq
and in particular,
\beq
N^{\phi'}(r_+) = 0.
\label{a13f}
\eeq
The identifications \rref{a11}, expressed in terms of the ``usual
Schwarzschild angle'' $\phi'$, thus lead to a vanishing shift at the
horizon.

\subsection{A Conical Singularity with a Helical Twist}

As in the static case, it is straightforward to extend the rotating black
hole solution to one admitting a conical singularity along the $z$ axis.
A slight generalization is now natural.  Recall that for the static black
hole, the periodic coordinates $\ln\!R$ and $\theta$ parametrized the
two circumferences of the torus $\chi=\hbox{\em const}$.  The natural
generalization of the periodicity \rref{a7a} in $\theta$ therefore
includes an associated twist in the $\ln\!R$ direction; that is,
\beq
(R,\theta,\chi) \sim
  (e^\Sigma R, \theta + \Theta, \chi) .
\label{a14a}
\eeq
As before, the nonstandard periodicity in $\theta$ indicates the presence
of a conical singularity along the $z$ axis, while the shift in $\ln R$
represents a simultaneous ``radial'' twist, resembling the time helical
structure of the spacetime around a spinning particle in 2+1 dimensions
\cite{DJtH}.

Note that the identifications \rref{a12} are not altered by the
introduction of this singularity, so equation \rref{a12a} is maintained.
The Schwarzschild coordinate periods $\Phi$ and $\beta$ appearing in
\rref{a12b}, on the other hand, now become
\beq
\beta = {\ell^2\over r_+^2 - r_-^2}\bigl( -|r_-|\Sigma + r_+\Theta\bigr)
       ,\qquad
\Phi = {\ell\over r_+^2 - r_-^2}\bigl( r_+\Sigma + |r_-|\Theta\bigr) .
\label{a14b}
\eeq
Correspondingly, it is now necessary to replace \rref{a13b} by
\beq
\phi' = \phi + {r_+\Sigma + |r_-|\Theta\over |r_-|\Sigma - r_+\Theta}\cdot
  {\tau\over\ell} ,
\label{a14c}
\eeq
which generates a shift
\beq
N^{\phi'}(r) = -{r_+\Sigma + |r_-|\Theta\over\ell(|r_-|\Sigma - r_+\Theta)}
  - {r_+|r_-|\over\ell r^2} .
\label{a14d}
\eeq
In particular, it follows that
\beq
\Sigma=0 \Leftrightarrow N^{\phi'}(r_+) = 0 ,
\label{a14e}
\eeq
and, once this is taken into account,
\beq
\Theta = 2\pi \Leftrightarrow \beta = {2\pi r_+\ell^2\over r_+^2 - r_-^2} .
\label{a14f}
\eeq
We shall see in section 2 how the conditions \rref{a14e} and \rref{a14f}
reappear in a very different approach.

\subsection{Isometries and Holonomies}

It is well known that the full isometry group of the hyperbolic metric
\rref{a4} is the group SL(2,$\IC$), the universal covering group of
the Lorentz group.  It is useful to express the isometry \rref{a12}
explicitly as an element of this group.  The action of SL($2,\IC$) is
most easily expressed in the language of of quaternions \cite{Bear}:
if we write the coordinates $x$, $y$, and $z$ as a quaternion
\beq
q = x + y{\bf i} + z{\bf j} ,
\label{a14}
\eeq
then a matrix
\beq
\pmatrix{a&b\cr c&d\cr}\in\hbox{SL(2,$\IC$)} ,\qquad a,b,c,d \in \IC
\label{a15}
\eeq
acts by
\beq
q \mapsto (aq+b)(cq+d)^{-1} .
\label{a16}
\eeq
It is easy to show that the identifications \rref{a12} are represented
by the matrix
\beq
H = \pmatrix{e^{\pi(r_+ + i|r_-|)/\ell} &0\cr
             0& e^{-\pi(r_+ + i|r_-|)/\ell}\cr} .
\label{a17}
\eeq
$H$ itself is coordinate-dependent, but its trace gives an invariant
characterization of the geometry.

Experience from elsewhere in (2+1)-dimensional gravity \cite{Witten,Carlip}
has taught us that it is often useful to look at the theory in first-order
form, with variables consisting of a triad $e^a = e^a{}_\mu dx^\mu$ and a
spin connection $\omega^a = {1\over2}\epsilon^{abc}\omega_{bc\mu} dx^\mu$.
For a Euclidean spacetime with a negative cosmological constant, these
two one-forms can be combined to form a single SL(2,$\IC$) connection
\beq
A^a = \omega^a + {i\over\ell}e^a ,
\label{a18}
\eeq
and Witten has shown \cite{Witten} that the ordinary Einstein action then
reduces to the Chern-Simons action
\beq
I = {i\ell\over 64\pi G} \int_M d^3\!x\, \epsilon^{ijk}\!
  \left[ A_i^a(\partial_j A_{ka} - \partial_k A_{ja})
  + {2\over3}\epsilon_{abc}A_i^aA_j^bA_k^c\right] + \hbox{c.c.}
\label{a19}
\eeq
In particular, if we start with the hyperbolic metric \rref{a4} in the
coordinates \rref{a6}, the triad can be chosen to be
\beq
  e^1 = {\ell\over\sin\chi}{dR\over R} , \quad
  e^2 = {\ell\over\sin\chi}d\chi , \quad
  e^3 = \ell\cot\chi d\theta
\label{a20}
\eeq
with a corresponding spin connection
\beq
  \omega^1 = -{1\over\sin\chi}d\theta , \quad
  \omega^2 = 0 , \quad
  \omega^3 = \cot\chi {dR\over R} .
\label{a21}
\eeq
It is then straightforward to show that $A^a$ is flat, as required by
the Chern-Simons field equations.  We stress that this flatness
does not imply triviality---although its curvature vanishes, $A^a$ has
a nonvanishing holonomy around the noncontractible closed curve
\beq
\gamma_A: s\mapsto \left( e^{2\pi r_+ s/\ell},
  \theta_0 + {2\pi|r_-|\over\ell}s, \chi_0 \right) ,\quad s\in[0,1]
\label{a22}
\eeq
connecting the inner and outer hemispheres of figure 1.  The value of
this holonomy can be computed directly from the connection, but it is
more easily understood geometrically: up to conjugation by a rather
complicated coordinate-dependent matrix, it is precisely the matrix
\rref{a17} describing the identifications that characterize the black
hole.  Cangemi et al.\ \cite{Cangemi} have found an analogous result
for the Lorentzian black hole, but a poor coordinate choice made their
computations rather difficult; see also \cite{BHTZ} and \cite{Vaz}.

It is perhaps worth emphasizing this relationship between identifications
and holonomies.  The direct solution of the Einstein field equations leads
to the Euclidean black hole metric \rref{a9a}.  On the other hand, we know
that any vacuum solution in three dimensions with $\Lambda<0$ must have
constant negative curvature, and ought therefore to be obtainable from
$\IH^3$ by suitable identifications.  The Chern-Simons formulation provides
the bridge between these two pictures: the required identifications
of $\IH^3$ are precisely the holonomies of the connection \rref{a18}.
Mathematically, this correspondence comes from the relationship between
flat connections and geometric structures, a subject of considerable
research in the past few years \cite{geom}.

We conclude this section with a brief discussion of the large
diffeomorphisms.  Since the Euclidean black hole is topologically a
solid torus, we should expect to find a one-parameter family of large
diffeomorphisms, corresponding to Dehn twists (twists by multiples of
$2\pi$) around the contractible circumference \cite{Dehn}.  These
diffeomorphisms are evident in \rref{a12}: the identifications that
determine the geometry are unchanged by the replacement
\beq
r_+\rightarrow r_+, \quad |r_-| \rightarrow |r_-| + n\ell
\label{a23}
\eeq
for any integer $n$.  We thus find a rather strange symmetry, under
which solutions with different masses and angular momenta are related
by large diffeomorphisms.  We do not yet understand the physical
significance of this invariance.

\subsection{Holonomies as Canonical Variables}

We now turn to a discussion of the dynamical variables of the system
and the corresponding Poisson brackets.  In the Chern-Simons
description of (2+1)-dimensional gravity, the fundamental physical
observables are the holonomies of the connection $A^a$, defined by
\rref{a18}, around curves $\gamma$ in $M$.  It is almost possible
to express these holonomies as functions of homotopy classes $[\gamma]$
alone: under a smooth deformation of $\gamma$, the holonomy $H(\gamma)$
of a flat connection is invariant up to overall conjugation.  In particular,
the traces $\Tr H(\gamma)$ are homotopy-invariant, and if $\gamma$ is
contractible, the entire holonomy matrix is the identity.

The Poisson algebra of the holonomies $H(\gamma)$ has been analyzed in
detail by Nelson and Regge \cite{NR,NR2}.  The holonomies around two
curves $\gamma_1$ and $\gamma_2$ have nonvanishing Poisson brackets only
when the projections of $\gamma_1$ and $\gamma_2$ onto a surface of
constant time intersect and cannot be separated by a smooth deformation.
The resulting algebra is simplest when spacetime has the topology
$\IR\!\times\!\Sigma$, with $\Sigma$ a closed genus $g$ surface.
In that case, the homotopy classes $[\gamma]$ automatically come in $2g$
intersecting pairs, the archetype being the two independent circumferences
of a torus.  It may be shown that the resulting Poisson brackets give rise
to a natural symplectic structure on the space of observables, thus
permitting a simple Hamiltonian formulation of the theory \cite{Goldman}.

For the Euclidean black hole, on the other hand, there is only one
nontrivial homotopy class, $[\gamma_A]$, and only one nontrivial holonomy.
It would thus seem that there is no classical symplectic structure, and
no starting point for this approach to quantization.

Fortunately, this is not quite true.  Consider first a section of the
geometry \rref{a2} lying between times $\tau_1$ and $\tau_2$, as illustrated
in upper half-space coordinates in figure 2.  (The figure shows
the $J=0$ case, but the generalization to nonzero spin is straightforward.)
In addition to the noncontractible path $\gamma_A$ discussed above, there
is now a nontrivial path $\gamma_B$ joining the constant time surfaces
$\tau=\tau_1$ and $\tau=\tau_2$.  The curves $\gamma_A$ and $\gamma_B$
are linked, and their holonomies therefore have nontrivial Poisson brackets.

More spcifically, one can evaluate the Poisson brackets by ``radial
quantization,'' treating the radial Schwarzschild coordinate $r$---or
the corresponding coordinate $\chi$ in the upper half-space
representation---as time.  It is easy to check that for a curve
$(R(s),\theta(s))$ on a surface of constant $\chi$, the holonomy of
the connection formed from \rref{a20}--\rref{a21} is
\beq
H = \pmatrix{\cos w + i\cot\chi\sin w&-{\csc\chi\sin w}\cr
  -{(\cos^2\chi-\sin^2\chi)\csc\chi}\sin w&\cos w - i\cot\chi\sin w\cr} .
\label{b1}
\eeq
with $w = \theta - i\ln R$.  Using the identifications \rref{a12}, we obtain
for the curve $\gamma_A$
\beq
\Tr H(\gamma_A) = 2\cosh{2\pi\over\ell} \left( r_+ + i|r_-| \right) ,
\label{b2}
\eeq
while for a curve $\gamma_B$ connecting $(R_1,\theta_1)$ and
$(R_2 = e^\Sigma R_1,\theta_2 = \theta_1 + \Theta)$,
\beq
\Tr H(\gamma_B) = 2\cosh\left( \Sigma + i\Theta \right) .
\label{b3}
\eeq
The Poisson brackets can then be read off from equation (5.2) of \cite{NR2};
with factors of $G$ restored, we obtain
\begin{eqnarray}
\left\{r_+, \Theta\right\} &=& \left\{ |r_-|, \Sigma\right\} =
   4G \nonumber\\
\left\{ |r_-|, \Theta\right\} &=& \left\{r_+, \Sigma\right\} = 0 .
\label{b4}
\end{eqnarray}

If we now return to the complete, periodic Euclidean black hole---figure
1 rather than figure 2---we see that $\Theta$ and $\Sigma$ are precisely
the deficit angle at the horizon and the associated helical twist described
in section 1.2.  It was shown in \cite{BTZ2} that the horizon area is
canonically conjugate to the opening angle for a black hole in any number
of dimensions.  Since the horizon area in 2+1 dimensions is proportional
to $r_+$, this agrees with \rref{b4}.

Of course, the complete, classical vacuum solution requires that there be
no conical singularity.  But the Poisson brackets \rref{b4} show that such
a requirement is inconsistent in the quantum theory, where the deficit angle
and the horizon radius are complimentary observables.  We shall see in the
next section that the same formulation arises naturally in a minisuperspace
approach to the black hole.

\section{Minisuperspace and Radial Quantization}
\addtocounter{footnote}{-1}

We now turn to a Hamiltonian approach to the Euclidean black hole, and
develop a minisuperspace model that will provide further insight into
the system.  We shall see that many of the conclusions of section
1---including the role of the horizon radius and the deficit angle as
conjugate variables---may be duplicated in this approach, in a context
that is directly generalizable to 3+1 dimensions.

\subsection{The Action and Boundary Terms}

As discussed in reference \cite{BTZ2}, the Euclidean action for a black
hole may be taken to be\footnote{Similar results have been obtained in
references \cite{Brown,Hay,Hay2} in slightly different contexts.}
\beq
I = {1\over4G}(\hbox{area of horizon}) + I_{\hbox{\scriptsize\it can}}
  + B_\infty ,
\label{c1}
\eeq
where $I_{\hbox{\scriptsize\it can}}$ is the canonical (ADM) action
and $B_\infty$ is a local boundary term at large spatial distances whose
form depends on what is held fixed at infinity.  (The sign is such that
one path integrates $e^{+I}$.)  For the complete black hole spacetime,
the action \rref{c1} differs from the Hilbert action
\beq
I_H={1\over8\pi G}\left[{1\over2}\int\sqrt{g}\left(R+2\ell^{-2}\right)d^3x
  -\int_{\partial M}\sqrt{h}Kd^2x\right]
\label{c2}
\eeq
by another boundary term at infinity.  In order to agree with the
conventions used in \cite{BHTZ}, we shall set $G=1/8$ in this section.
We now specialize \rref{c1} to a class of fields that includes
our rotating black hole, but in which all fluctuations in azimuthal
angle and time are frozen.  Thus we admit all metrics of the form
\beq
ds^2 = \beta^2(r)f^2(r)d\tilde\tau^2 + f^{-2}(r)dr^2 +
  r^2\left(d\phi' + \tilde N^{\phi'}(r)d\tilde\tau\right)^2 ,\qquad
  r_+\le r <\infty,\  0\le\tilde\tau\le1
\label{c3}
\eeq
with
\beq
f^2(r_+) = 0
\label{c4}
\eeq
and
\beq
f^2(r) - \left({r\over\ell}\right)^2 {\rightarrow}  -M \quad\hbox{as}\
 {r\rightarrow\infty}.
\label{c5}
\eeq
We have replaced the lapse function $N^\perp(r)$ with the ``Killing
lapse''
\beq
\beta(r) = f^{-1}(r)N^\perp(r)
\label{c6}
\eeq
for later convenience, and have denoted the azimuthal angle by $\phi'$ to
agree with the notation of section 1.

The canonical Euclidean action has the form
\beq
I_{\hbox{\scriptsize\it can}}
  = \int d\tilde\tau d^2x\,
  \left( \pi^{ij}{\partial g_{ij}\over\partial\tilde\tau}
  - N^\perp{\cal H}_\perp - N^i{\cal H}_i\right) ,
\label{c7}
\eeq
and specialized to the minisuperspace \rref{c3}, it becomes \cite{BHTZ}
\beq
I_{\hbox{\scriptsize\it can}}
  = -\int_{r_+}^\infty dr\, \left\{ \beta(r) \left[
  (f^2)'(r) - {p^2(r)\over 2r^3} - {2r\over\ell^2}\right]
  + \tilde N^{\phi'}(r)p'(r)\right\} ,
\label{c8}
\eeq
where $p(r)$ is the $r$-$\phi$ component of the gravitational momentum,
related to the extrinsic curvature, and the prime denotes differentiation
with respect to $r$.
We shall be interested in the action principle based on \rref{c1} with
boundary conditions that permit the existence of a unique classical
solution.  This is because we wish to investigate the classical and
semiclassical approximations to the path integral, which start with
the classical action $\bar I$ for an extremum with specified boundary
conditions.  We therefore fix the functions $\beta(r)$ and
$\tilde N^{\phi'}(r)$ at spatial infinity,
\beq
\beta(r)\rightarrow\beta(\infty) , \quad
  \tilde N^{\phi'}(r)\rightarrow \tilde N^{\phi'}(\infty)
  \quad\hbox{as}\quad {r\rightarrow\infty} ,
\label{c10}
\eeq
and we also fix
\beq
p(r_+) = p_+
\label{c11}
\eeq
to supplement \rref{c4}.  Equations \rref{c4}, \rref{c10}, and
\rref{c11} provide a complete set of boundary conditions for the
action principle, and the four numbers $r_+$, $p_+$, $\beta(\infty)$,
and $\tilde N^{\phi'}(\infty)$ fully determine the classical solution.
Since the ``momenta'' $\beta$ and $\tilde N^{\phi'}$ are fixed at the
upper end point, one must add to the action \rref{c7} a boundary term
$$-\beta(\infty)M - \tilde N^{\phi'}(\infty)J ,$$
where neither the parameter $M$ defined in \rref{c5} nor $J = -p(\infty)$
are held fixed.

When varied with respect to $f^2$, $p$, $\beta$, and $\tilde N^{\phi'}$
with these boundary conditions, the full action
\beq
I = 4\pi r_+ + I_{\hbox{\scriptsize\it can}}
  -\beta(\infty)M - \tilde N^{\phi'}(\infty)J
\label{c14}
\eeq
has an extremum when the following equations hold:
\begin{eqnarray}
&&(f^2)' - {p^2\over 2r^3} - {2r\over\ell^2} = 0 \nonumber\\
&& p' = 0 \\
&& \beta' = 0 \nonumber\\
&& (\tilde N^{\phi'})' + {p\beta\over r^3} = 0 .\nonumber
\label{c15}
\end{eqnarray}
The solution is given by
\begin{eqnarray}
&&f^2(r) = -\left({r_+^2\over\ell^2} - {p_+^2\over4r_+^2}\right)
  + {r^2\over\ell^2} - {p_+^2\over4r^2} \nonumber\\
&&p(r) = p_+ \\
&&\beta(r) = \beta(\infty) \nonumber\\
&&\tilde N^{\phi'}(r) = \tilde N^{\phi'}(\infty)
  + {p_+\beta(\infty)\over2r^2} .\nonumber
\label{c19}
\end{eqnarray}

\subsection{Canonical Variables}

The action \rref{c8} is in canonical form, with the radial coordinate
$r$ playing the role of time (``radial quantization'').\footnote{A
similar radial foliation has been considered by Brown et al.\ \cite{Brown}
in the context of black hole thermodynamics.}  It contains two canonical
pairs, $(\beta, f^2)$ and $(\tilde N^{\phi'},p)$, with equal $r$ brackets
given by
\beq
\left\{ \beta,f^2\right\} = \left\{\tilde N^{\phi'},p\right\} = 1 .
\label{c9}
\eeq
In section 1, on the other hand, we described the black hole in terms of
two different pairs of canonically conjugate variables, $(r_+,\Theta)$
and $(|r_-|,\Sigma)$.  We now further investigate the connection
between these pairs.

Let us begin by comparing the solution \rref{c19} to the metric
\rref{a9a}--\rref{a9b}, permitting the periodicities \rref{a14b} that
describe a general conical singularity.  It is easy to check that the
two metrics are identical if we set
\beq
\tilde\tau = \tau/\beta ,\qquad \tilde N^{\phi'} = \beta N^{\phi'}
\label{ca1}
\eeq
with
\begin{eqnarray}
&&p_+ = -J = -{2r_+|r_-|\over\ell} \nonumber\\
&&M = {r_+^2 + r_-^2\over\ell^2} \\
&&\beta(\infty) = \beta
  = {\ell^2\over r_+^2 - r_-^2}\bigl( -|r_-|\Sigma + r_+\Theta\bigr)
  \nonumber\\
&&\beta^{-1}\tilde N^{\phi'}(\infty) = N^{\phi'}(\infty) =
  -{r_+\Sigma + |r_-|\Theta\over\ell(|r_-|\Sigma - r_+\Theta)}.\nonumber
\label{ca2}
\end{eqnarray}

We can now compute the Poisson brackets $\{\beta, f^2\}$ and $\{\tilde
N^{\phi'},p\}$ directly from the brackets \rref{b4}.  We find that the
only nonvanishing equal $r$ brackets are
\beq
\{ \beta, f^2 \}
  = \left\{ {\ell^2\over r_+^2 - r_-^2} (-|r_-|\Sigma + r_+\Theta) ,
  {|r_-|^2 - r_+^2\over\ell^2} - {r_+^2|r_-|^2\over\ell^2r^2},\right\}
  = 1
\label{ca3}
\eeq
and
\beq
\{ \tilde N^{\phi'}, p_+ \}
  = \left\{ {\ell r_+|r_-|\over r^2(r_+^2-r_-^2)}(|r_-|\Sigma - r_+\Theta)
  + {\ell\over r_+^2-r_-^2}(r_+\Sigma + |r_-|\Theta) ,
  - {2r_+|r_-|\over\ell} \right\} = 1 ,
\label{ca4}
\eeq
in agreement with \rref{c9}.

We can thus view the Chern-Simons quantization of section 1 as a form
of ``covariant canonical quantization,'' that is, quantization of the
space of classical solutions \cite{covcan,Crn}.  The space of classical
solutions, which is isomorphic to the reduced phase space, is parametrized
by constants of motion $\{r_+,r_-,\Sigma,\Theta\}$, and the canonical
commutation relations of the full theory---or, in this case, the
minisuperspace model---are equivalent to the commutation relations
among these parameters.

An alternative derivation of the conjugacy of $r_+$ and $\Theta$ may
offer further insight.  In the canonical action \rref{c8}, $r$ clearly
plays the role of a ``time'' variable, with a corresponding ``Hamiltonian''
\beq
H_r(r) = -\beta(r)\left[ {p^2(r)\over 2r^3} + {2r\over\ell^2}\right] .
\label{ca5}
\eeq
It is then a standard result that the variable conjugate to $r_+$ is
$-H_r(r_+)$, which by equation \rref{c15} is equal to $(\beta f^2)'(r_+)$.
But from \rref{a7b}, this is just the opening angle, or rather $2\Theta$.
The $r_+$ dependence of the remaining terms in \rref{c14} gives an
additional contribution of $-4\pi$ to the momentum conjugate to
$r_+$; combining the two contributions, we find
\beq
P_{r_+} = -2(2\pi - \Theta) ,
\label{c24}
\eeq
in accord with the general discussion of \cite{BTZ2} and in agreement
with \rref{a14e}.  The extra $4\pi$ does not contribute to the Poisson
brackets, but it will be important in the determination of the partition
function in the next section.

\subsection{Partition Function}

We now turn to the computation of the thermal partition function of the
(2+1)-dimensional black hole.  The partition function may be obtained as
a trace of the Euclidean propagation amplitude, which is derived in turn
from the path integral for the wedge-shaped manifold $\tau_1\le\tau\le\tau_2$
of figure 2.  In this section, we shall concentrate on the classical
approximation, in which the propagation amplitude is expressed as the
exponential of the classical action $\bar I$; the first quantum correction
will be described in the next section.

The Euclidean propagation amplitude
$$K[{}^{(2)}{\cal G}_2,{}^{(2)}{\cal G}_1;r_+,p_+;
  \beta(\infty),\tilde N^{\phi'}(\infty)]$$
depends on the two-geometries of the slices $\tau=\tau_1$ and $\tau=
\tau_2$; the horizon geometry, which is  determined by $r_+$ and $p_+$;
and the asymptotic geometry, characterized by $\beta(\infty)$ and
$\tilde N^{\phi'}(\infty)$.  Upon performing a suitable Laplace transform,
the latter two parameters may be replaced by the mass $M$ and the angular
momentum $J$.  It is to be emphasized that the asymptotic constants $M$
and $J$ are to be treated as independent parameters; they are determined
in terms of $r_+$ and $p_+$ only ``on shell.''

To obtain the partition function, we must take the trace of $K$ over
the initial and final geometries, including the horizon geometries.
The metric in our minisuperspace model is $\tau$-independent, so
${}^{(2)}{\cal G}_1$ and ${}^{(2)}{\cal G}_2$ are automatically equal,
and the trace reduces to an integral over $r_+$ and $p_+$.  For the
full path integral, one must be careful about the contour of integration
and the measure---see \cite{Whiting1,Whiting2} for a related discussion
in a (3+1)-dimensional setting---but for the classical approximation,
this integration amounts to extremizing the action with respect to $r_+$
and $p_+$, to obtain the grand canonical partition function with the
temperature $\beta(\infty)$ and the rotational chemical potential
\beq
\mu = (\beta^{-1}\tilde N^{\phi'}_{\hbox{\scriptsize Lor}})(\infty)
\label{c22c}
\eeq
held fixed.  (Note that $\tilde N^{\phi'}(\infty)J$ is equal to
$(-i \tilde N^{\phi'}_{\hbox{\scriptsize Lor}}(\infty))\cdot
(-iJ_{\hbox{\scriptsize Lor}}) = -\beta\mu J_{\hbox{\scriptsize Lor}}$.)

Extremizing with respect to $r_+$ and $p_+$ amounts to setting their
canonical conjugates equal to zero.  The variable conjugate to $p_+$ is
$\tilde N^{\phi'}(r_+)$, so we find
\beq
\tilde N^{\phi'}(r_+) = 0 ,
\label{c23}
\eeq
in agreement with \rref{a14e}.  The variable conjugate to $r_+$ is
$P_{r_+} = -2(2\pi-\Theta)$, so $\Theta=2\pi$ at the extremum, again in
accord with the results of section 1.  Moreover, the canonical action
$I_{\hbox{\scriptsize\it can}}$ vanishes when the equations of motion
\rref{c15} are satisfied.  We therefore obtain an extremal action of
\beq
{\bar I}[\beta,\mu] = 4\pi r_+ - \beta M - \tilde N^{\phi'}(\infty)J
\label{c31}
\eeq
with $M$, $J$, and $r_+$ expressed in terms of $\beta$ and $\tilde
N^{\phi'}(\infty)$ through \rref{ca2} with $\Sigma=0$ and $\Theta=2\pi$.
The classical approximation to the partition function is therefore
\beq
Z(\beta,\mu) = e^{\bar I} ,
\label{c32}
\eeq
and the entropy in this approximation is
\beq
S = 4\pi r_+ ,
\label{c33}
\eeq
or reinstating the universal constants,
\beq
S = {2\pi r_+\over4\hbar G} .
\label{c34}
\eeq
Note that in conventional units ($\hbar=G=1$), the entropy is just
a quarter of the horizon size, as expected for Einstein's theory in any
spacetime dimension \cite{BTZ2}.

\subsection{First Quantum Correction}

The results of the previous section are essentially those of a tree-level
approximation.  In particular, the exponent appearing in equation \rref{c32}
is the classical action of the Euclidean black hole.  By returning to the
Chern-Simons formulation, however, we can easily obtain the first quantum
(``one-loop'') correction.

As we observed in section 1.1, the Euclidean black hole is topologically
a solid torus.  For such a topology, the one-loop contribution to the
Chern-Simons path integral has already been worked out, albeit in a rather
different context, in appendix 3 of reference \cite{Carlip2}.  This
contribution---essentially the Van Vleck-Morette determinant---takes
a very simple form, depending only on the holonomy \rref{a17}: the
prefactor is simply
\beq
\Delta =
 4\pi\left( \cosh {2\pi r_+\over\ell} - \cos {2\pi|r_-|\over\ell} \right) .
\label{e1}
\eeq
For $r_+/\ell$ large, this becomes
\beq
\Delta \approx 2\pi e^{2\pi r_+/\ell},
\label{e2}
\eeq
so the entropy \rref{c34} is corrected to read
\beq
S = {2\pi r_+\over\ell}\left( {\ell\over4\hbar G} + 1\right) .
\label{e3}
\eeq
Note that while the ``classical'' term involves the area expressed in
Planck units, the first quantum correction depends instead on the scale
$\ell$ set by the cosmological constant, and involves neither $\hbar$
nor $G$.

In some approaches to quantum gravity, the partition function also
receives contributions from other topologies.  These contributions may
be quite large, and in some cases they dominate the path integral
\cite{Carlip2,Carlip3}.  For the Euclidean black hole, however, this is
not the case.  It will be shown elsewhere \cite{inprep} that the only
complete extrema of the Euclidean action with the {\em asymptotic} geometry
of the black hole are the black hole itself and the ``hot empty space''
solution obtained by identifying
\beq
(x,y,z) \sim (x+1,y,z) \sim (x+\tau_1,y+\tau_2,z)
\label{e4}
\eeq
in the metric \rref{a4}.

\section{Steps Towards Black Hole Statistical Mechanics}

We have now seen that the thermal properties of the (2+1)-dimensional
black hole are intimately related to the existence of new degrees of
freedom associated with a conical singularity at the Euclidean
event horizon.  It is shown in reference \cite{offshell} that
a similar conclusion holds for the black hole in any number of spacetime
dimensions.  The off-shell Euclidean black hole in $d$ dimensions has the
topology $\IR^2\!\times\!S^{d-2}$, and just as in the three-dimensional
case, one must generically allow a conical singularity in the $\IR^2$
plane.  As in the previous section, such a singularity leads to a term
$(2\pi - \Theta)\delta A$ in the variation of the Hilbert action,
where $A$ is now the area (volume if $d>4$) of the $(d-2)$-sphere
at the horizon; it is shown in \cite{BTZ2} that this term has its
geometrical origin in the dimensional continuation of the two-dimensional
Gauss-Bonnet theorem.  This boundary variation implies that the opening
angle $\Theta$ is canonically conjugate to the area of the horizon,
in agreement with \rref{b4} for 2+1 dimensions.

This is an attractive picture, but it does not yet provide a
``statistical mechanical'' explanation of the black hole entropy \rref{c34}.
Ideally, one would like to explain this entropy as a logarithm of the
number of macroscopically indistinguishable states.  Our analysis suggests
that these states ought to be associated with the conical singularity at
$r=r_+$, but a more detailed microscopic description seems to require a
full quantization of the black hole.

We do not yet know how to complete such a program.  We can, however,
point to several suggestive results:

\subsection{Counting Euclidean States}

One starting point for black hole statistical mechanics is the
Chern-Simons approach to three-dimensional gravity.  As discussed in
section 1.2, three-dimensional Euclidean gravity can be described by
the Chern-Simons action \rref{a19} with gauge group SL($2,\IC$).  The
quantization of such an action has been discussed by Witten \cite{Wit2}
and Hayashi \cite{Hayashi}.  In the special case that the coupling
constant ${\ell/8\hbar G}$ is an integer, Hayashi argues that the space
of states on a torus is isomorphic to two copies of the Hilbert space
of an SU($2$) Chern-Simons theory with coupling constant $k=\ell/8\hbar G$.
The group SU($2$) appears because SL($2,\IC$) is the complexification of
SU($2$); roughly speaking, the two SU($2$) gauge fields are the connection
$A^a$ of \rref{a18} and its complex conjugate, treated as independent
fields.

The states of an SU($2$) Chern-Simons theory can be created by inserting
a Wilson line carrying an SU($2$) representation of spin $j/2$ ($j = 1,
\dots,k$) at the core of a solid torus.  Since such Wilson lines are the
Chern-Simons analog of conical singularities \cite{Witten,Carlip4}, this
description closely resembles the analysis of section 1.  In this relatively
simple approach to quantization, however, there appears to be no evidence
for the exponentially rising density of states needed to explain black
hole entropy; it may be shown that one only obtains, roughly speaking,
a set of states evenly spaced in $r_+$.

This straightforward version of the Chern-Simons quantization can be
replaced, however, by one in which the role of the horizon becomes more
fundamental.  For the Lorentzian black hole, the horizon is not a single
line, but is rather a boundary beween the interior and exterior regions.
To mimic this feature in the Euclidean solution, we can remove a small
cylinder around the horizon $r=r_+$, obtaining a Euclidean version of
the ``stretched horizon'' \cite{Thorne,Suss}.

The introduction of such a boundary dramatically changes the SL($2,\IC$)
Chern-Simons theory: the Chern-Simons action now induces a dynamical
chiral Wess-Zumino-Witten action on the stretched horizon, whose states
represent genuine horizon degrees of freedom.  For a noncompact group
like SL($2,\IC$), the structure of the resulting WZW Hilbert space is
poorly understood.  But we can again look for analogies with the SU($2$)
theory, as suggested by Witten in the last section of reference \cite{Wit2}.

An SU(2) WZW model has an infinite number of states, but only finitely many
of these occur for any given eigenvalue of the Virasoro operator $L_0$.
Now, $L_0$ can be interpreted as a generator of time translations, so we
should expect its eigenvalues to be proportional to the mass of the black
hole.  For large values of $L_0$, it is known that the number of states
increases exponentially with $L_0^{1/2}$ \cite{Kac}.  Since the mass of
the three-dimensional black hole is proportional to $r_+^2$, this gives at
least the right qualitative dependence of the number of states on the
horizon size.  The exponential dependence of the density of states on
$L_0^{1/2}$ is a generic feature of Ka{\v c}-Moody algebras, at least for
those based on compact Lie groups \cite{Kac}, so this result may not be
too sensitive to our use of SU($2$) as a stand-in for the correct gauge
group.

\subsection{Topological Field Theory and Four-Dimensional Black Holes}

The results of the last section appear to be peculiar to the
three-dimensional black hole.  A generalization to four dimensions lies
beyond the scope of this paper.  We would, however, like to make two
brief observations.

First, the geometry of the Euclidean black hole is largely independent
of the dimension of spacetime.  The black hole in $d$ dimensions has
the topology $\IR^2\!\times\!S^{d-2}$, and the entropy comes quite
generically from the existence of a possible conical singularity in the
$\IR^2$ plane \cite{offshell,BTZ2}.  The exponential form of the entropy can
be interpreted as the existence of ``one degree of freedom on $\IR^2$
per unit horizon area.''  An understanding of the role of $\IR^2$
in any dimension---essentially a question of topological field theory
on a cone---might therefore lead to a microscopic interpretation of
black hole entropy in all dimensions.

Second, although the Chern-Simons formulation is unique to three dimensions,
the meaning of the WZW  degrees of freedom can be generalized.  The
SL($2,\IC$) Wess-Zumino-Witten theory on the horizon arises because it
is not consistent to gauge-fix all of the ``gauge'' degrees of freedom;
the horizon dynamics is that of would-be gauge transformations that are
forced to become dynamical \cite{EMSS,Liou}.  But SL($2,\IC$) gauge
transformations in the Chern-Simons formulation are equivalent to
diffeomorphisms and local Lorentz transformations in the metric formulation.
We therefore suggest that an understanding of black hole entropy is likely
to require a careful analysis of the process of gauge-fixing at the horizon.
We hope to return to these questions elsewhere.

\vspace{1ex}
\begin{flushleft}
\large\bf Acknowledgements
\end{flushleft}

We would like to thank M\'aximo Ba\~nados, Frank Wilczek, and Jorge Zanelli
for enlightening discussions.  This work was partially supported by grants
0862/91 and 193.1910/93 from FONDECYT (Chile), by institutional support to
the Centro de Estudios Cientificos de Santiago provided by SAREC (Sweden)
and a group of Chilean private companies (COPEC, CMPC, ENERSIS, CGEI).
This research was also sponsored by CAP, IBM and Xerox de Chile.  S.C.\ was
supported in part by U.S.\ Department of Energy grant DE-FG03-91ER40674 and
National Science Foundation Young Investigator award PHY-93-57203.

\vspace{1ex}
\begin{flushleft}
\large\bf Figure Captions
\end{flushleft}
\begin{enumerate}
\item The Euclidean black hole is obtained by identifying the inner
and outer hemispheres of this figure along radial lines such as $L$.
The circle $\gamma_A$ is the segment of $L$ between the two hemispheres,
whose endpoints are identified.  (The outer hemisphere in this figure
has been cut open to show the inner hemisphere.)
\item Transition amplitudes are obtained from a section of the black
hole between two constant $\tau$ surfaces.  Each boundary of constant
$\tau$ is an annulus, with an inner circumference at $r_+$ (the $z$
axis) and an outer circumference at infinity (the intersection with
the $x$-$y$ plane).
\end{enumerate}


\begin{thebibliography}{99}
\bibitem{Bekenstein} J.\ D.\ Bekenstein, \PRD{7} (1973) 2333.
\bibitem{Hawking} S.\ W.\ Hawking, {\sl Nature} {\bf 248} (1974) 30.
\bibitem{Page} D.\ Page, ``Black Hole Information,'' Alberta preprint
 ALBERTA-THY-23-93 (to appear in {\sl Proc.\ of the Fifth Canadian
 Conference on General Relativity and Relativistic Astrophysics}).
\bibitem{BrownYork} J.\ D.\ Brown and J.\ W.\ York, \PRD{47} (1993) 1420.
\bibitem{BTZ} M.\ Ba\~nados, C.\ Teitelboim, and J.\ Zanelli, \PRL{69}
 (1992) 1849.
\bibitem{BHTZ} M.\ Ba\~nados, M.\ Henneaux, C.\ Teitelboim, and J.\ Zanelli,
 \PRD{48} (1992) 1506.
\bibitem{offshell} S.\ Carlip and C.\ Teitelboim, ``The Off-Shell Black
 Hole,'' IAS preprint IASSNS-HEP-93/84 and Davis preprint UCD-93-34 (1993).
\bibitem{DJtH} S.\ Deser, R.\ Jackiw, and G.\ 't Hooft, \Ann{152} (1984)
 220.
\bibitem{Bear} See, for example, A.\ F.\ Beardon, in {\sl Discrete Groups
 and Automorphic Functions}, ed.\ W.\ J.\ Harvey (Academic Press, N.Y.,
 1977).
\bibitem{Witten} E.\ Witten, \NPB{323} (1989) 113.
\bibitem{Carlip} S.\ Carlip, ``Six Ways to Quantize (2+1)-Dimensional
 Gravity, Davis preprint UCD-93-15 (to appear in {\sl Proc.\ of the Fifth
 Canadian Conference on General Relativity and Relativistic Astrophysics}).
\bibitem{Cangemi} D.\ Cangemi, M.\ Leblanc, and R.\ B.\ Mann, \PRD{48}
 (1993) 3606.
\bibitem{Vaz} C.\ Vaz and L.\ Witten, ``Wilson Loops and Black Holes
 in 2+1 Dimensions,'' Universidade do Algarve preprint UATP-93/06.
\bibitem{geom} See W.\ M.\ Goldman, in {\sl Geometry of Group
 Representations}, eds.\ W.\ M.\ Goldman and A.\ R.\ Magid (Amer.\
 Math.\ Soc., Providence, 1988).
\bibitem{Dehn} See, for example, J.\ S.\ Birman, in {\sl Discrete Groups
 and Automorphic Functions}, ed.\ W.\ J.\ Harvey (Academic Press, N.Y.,
 1977).
\bibitem{NR} J.\ E.\ Nelson and T.\ Regge, \NPB{328} (1989) 190.
\bibitem{NR2} J.\ E.\ Nelson, T.\ Regge, and F.\ Zertuche, \NPB{339} (1990)
 516.
\bibitem{Goldman} The mathematics of this symplectic structure is discussed
  by W.\ M.\ Goldman, {\sl Invent.\ Math.} {\bf 85} (1986) 263.
\bibitem{BTZ2} M.\ Ba\~nados, C.\ Teitelboim, and J.\ Zanelli, \PRL{72}
 (1994) 957.
\bibitem{Brown} J.\ D.\ Brown et al., \CQG{7} (1990) 1433.
\bibitem{Hay} G.\ Hayward, \PRD{47} (1993) 3275.
\bibitem{Hay2} G.\ Hayward and J.\ Louko, \PRD{42} (1990) 4032.
\bibitem{covcan} A.\ Ashtekar and A.\ Magnon, {\sl Proc.\ Roy.\ Soc.\
 (London)} {\bf A346} (1975) 375.
\bibitem{Crn} C.\ Crnkovic and E.\ Witten, in {\sl Three Hundred
 Years of Gravity}, eds.\ S.~W.\ Hawking and W.\ Israel (Cambridge
 University Press, Cambridge, 1987).
\bibitem{Whiting1} J.\ Louko and B.\ F.\ Whiting, \CQG{9} (1992) 457.
\bibitem{Whiting2} J.\ Melmed and B.\ F.\ Whiting, \PRD{49} (1994) 907.
\bibitem{Carlip2} S.\ Carlip, \PRD{46} (1992) 4387.
\bibitem{Carlip3} S.\ Carlip, \CQG{10} (1993) 207.
\bibitem{inprep} S.\ Carlip, in preparation.
\bibitem{Wit2} E.\ Witten, \CMP{137} (1991) 29.
\bibitem{Hayashi} N.\ Hayashi, {\sl Prog.\ Theor.\ Phys.\ Suppl.} {\bf 114}
 (1993) 125.
\bibitem{Carlip4} S.\ Carlip, \NPB{324} (1989) 106.
\bibitem{Thorne} K.\ S.\ Thorne, R.\ H.\ Price, and D.\ A.\ Macdonald,
 {\sl Black Holes: The Membrane Paradigm} (Yale University Press, 1986).
\bibitem{Suss} L.\ Susskind L.\ Thorlacius, and J.\ Uglum, ``The Stretched
 Horizon and Black Hole Complementarity,'' Stanford preprint SU-ITP-93-15
 (1993).
\bibitem{Kac} V.\ G.\ Ka{\v c} and D.\ H.\ Peterson, {\sl Adv.\ in Math.}
 {\bf 53} (1984) 125, section 4.7.
\bibitem{EMSS}  S.\ Elitzur et al., \NPB{326} (1989) 108.
\bibitem{Liou} S.\ Carlip, \NPB{362} (1991) 111.
\end{thebibliography}
\end{document}